\documentclass[prb,showpacs,amsfonts,eqsecnum,preprint,aps,byrevtex]{revtex4}
\usepackage{bm}

\newcommand{\ko}{K_0(x,t;x',t')}
\newcommand{\kop}{K^{(+)}_0(x,t;x',t')}
\newcommand{\kopy}{K^{(+)}_0(x,t;y,\theta)}

\newcommand{\dkomy}{\frac{\partial K^{(-)}_0}{\partial n_-(y)}(x,t;y,\theta)}
\newcommand{\dkop}{\frac{\partial K^{(+)}_0}{\partial n_+(x)}(x,t;x',t')}
\newcommand{\dkom}{\frac{\partial K^{(-)}_0}{\partial n_-(x)}(x,t;x',t')}
\newcommand{\kom}{K^{(-)}_0(x,t;x',t')}
\newcommand{\kpp}{K_{++}(x,t;x',t')}
\newcommand{\kpm}{K_{+-}(x,t;x',t')}
\newcommand{\kmm}{K_{--}(x,t;x',t')}
\newcommand{\kmp}{K_{-+}(x,t;x',t')}
\newcommand{\npp}{\nu_{++}(y,\theta;x',t')}
\newcommand{\npm}{\nu_{+-}(y,\theta;x',t')}
\newcommand{\nmm}{\nu_{--}(y,\theta;x',t')}
\newcommand{\nmp}{\nu_{-+}(y,\theta;x',t')}
\newcommand{\dt}{\int^t_{t'}\!\!d\theta\int_S \!\!dS_y\,}
\newcommand{\dto}{\int^t_{0}\!\!d\theta\int_S \!\!dS_y\,}
\begin{document}
\title{Integral
equations for heat kernel in compound media}
\author{I.~G.~Pirozhenko}
\affiliation{Bogoliubov Laboratory of Theoretical Physics, Joint Institute for
Nuclear Research, 141980 Dubna, Russia} \email{pirozhen@thsun1.jinr.ru}
\email{nestr@thsun1.jinr.ru}
\author{V.~V.~Nesterenko}
\affiliation{Bogoliubov Laboratory of Theoretical Physics, Joint Institute for
Nuclear Research, 141980 Dubna, Russia} \email{nestr@thsun1.jinr.ru}
\author{M.~Bordag}
\affiliation{Institute for Theoretical Physics, University of Leipzig,
Augustusplatz 10/11, 04109 Leipzig, Germany}
\email{Michael.Bordag@itp.uni-leipzig.de}
\date{\today}
\begin{abstract}
By making use of the potentials of the heat conduction equation the
integral equations are derived which determine the heat kernel for
the Laplace operator $-a^2\Delta$ in the case of compound media. In
each of the media the parameter $a^2$ acquires a certain constant
value. At the interface of the media the conditions are imposed which
demand the continuity of the `temperature' and the `heat flows'. The
integration in the equations is spread out only over the interface of
the media. As a result the dimension of the initial problem is
reduced by 1. The perturbation series for the integral equations
derived are nothing else as the multiple scattering expansions for
the relevant heat kernels. Thus a rigorous derivation of these
expansions is given. In the one dimensional case the integral
equations at hand are solved explicitly (Abel equations) and the
exact expressions for the regarding heat kernels are obtained for
diverse matching conditions. Derivation of the asymptotic expansion
of the integrated heat kernel for a compound media is considered by
making use of the perturbation series for the integral equations
obtained. The method proposed is also applicable to the
configurations when the same medium is divided, by a smooth compact
surface, into internal and external regions, or when only the region
inside (or outside) this surface is considered with appropriate
boundary conditions.
\end{abstract}
\maketitle
\section{Introduction}
The heat kernel technique\cite{DW,Bordag,Kirsten,Vassilevich,Od} is
widely used for constructing the quantum field theory in
gravitational background and with allowance for nontrivial boundary
conditions. Of a particular interest is the asymptotic expansion of
the heat kernel it terms of evolution parameter for its small values.
The coefficients of this  expansion pertain to divergences and
anomalies in the relevant quantum field theory models. Proceeding from this
one can develop the renormalization procedure needed.

For well posed spectral problems the heat kernel coefficients are expressed, in
an polynomial way, through the local geometric characteristics of the manifold
$D$ and its boundary $S$. Not only the contributions of $D$ and $S$ are
independent but the contributions of individual regions of $D$ and $S$ are also
additive.

The spectral problem is well posed for the goal of constructing the
heat kernel if the second oder elliptic differential operator in
question is close to the Laplace operator defined on a smooth
manifold with a smooth boundary, if any.

There are no universal methods for constructing the heat kernel and
its asymptotic expansion. The development of different approaches to
this problem is the subject of many works (see, for example, the
reviews\cite{DW,Bordag,Kirsten,Vassilevich,Od} and references
therein).

The initial definition of the heat kernel is the Green function of
the heat-conduction equation with an elliptic operator under study.
In many physical problems it is worth going from the differential
equation, defining the solution to be found or the relevant Green's
function, to the equivalent integral equation. In the dynamical
evolution problems the integral equations manifestly show the
reason-consequence relations governing the physical process under
study. Reducing the problem to the integral equation, as a rule,
allows one to develop the method of successive approximations
(perturbation theory).

Transforming the initial differential equation into the integral
form is in the general case a nontrivial problem. A special attention
should be paid here to incorporating the boundary conditions into the
integral equation. When constructing the integral equations governing
the Green's function of the heat equation we shall use the surface
potentials for this equation. The volume potential and the potentials
of single and double layers naturally arise in the theory of the
Laplace equation. In this case they are referred to as the Newtonian
(or electrostatic) potentials. This idea proved to be fruitful also
in studies of the Helmholtz equation describing, for example, steady
harmonic oscillations,  the wave equation, and the heat-conduction
equation. The potentials are particular solutions of these
homogeneous equations, and they  are constructed in a universal way
in terms of the fundamental (or elementary) solution of the initial
equation. The potential technique has turned out to be effective both
for consideration of the general properties of the equations under
study  (for example, the prove of the  solution uniqueness) and for
deriving particular solutions with given properties and for
obtaining the Green's functions.

This paper seeks to  demonstrate  the efficiency of using the heat
potentials when constructing the integral equations for the heat
kernel (Green functions) at first in the case of manifolds with
boundary. It is worth noting that integration in these equations
is spread out over the boundary only. As a result the
dimensionality of the initial problem is reduced by co-dimension
of the boundary $S$. Further  this approach is extended to the
compound media, where the principal part of the differential
operator has a discontinuity at the interface between different
media. A typical example here is the electrodynamics of continuous
media.\cite{Stratton} The velocity of light has, in the general
case, a jump discontinuity on the border between two  media with
different characteristics (for example, on the border between
dielectric and vacuum). In the both  media the Maxwell equations
are well defined, and at the interface the matching conditions (or
boundary conditions) should be satisfied. The concrete form of
these conditions is determined by the physical content of the
problem in question. In the same way the conduction of heat in
compound media is treated.\cite{Carslaw} As far as we are aware,
the heat kernel for compound media is not investigated
yet.\cite{Vassilevich}

The layout of the paper is as follows. In Sec.\  \ref{HP}  the
essentials of the potential theory are recalled first  for the
Laplace equation (Newtonian potentials) and then for the
heat-conduction equation (heat potentials). By making use of the heat
potentials the integral equation  for the Green function (heat
kernel) is derived for a compact region of Euclidean space bounded by
a smooth surface. The perturbative series for this equation is
developed which is nothing else as the multiple scattering expansion
for the heat kernel. Thus a rigorous derivation of this expansion is
presented.  The convenience to use here the Laplace transform is
shown. In Sec.\ \ref{CM} the integral equations are derived that
determine the heat kernel for compound media. The efficiency of the
approach proposed is demonstrated by deriving the $t$-small
asymptotics of the first three terms of the perturbation series for
the heat kernel in the case of compound media (Section~\ref{AE}). In
Sec.\ \ref{line} the heat kernel on an infinite line is constructed
in an exact form for diverse matching conditions. In the Conclusion
(Section \ref{Con}) the obtained results are briefly summarized and
the possibility of extending the approach proposed is discussed. In
Appendix \ref{appA} the general conditions at the interface are found
which result in self-adjoint boundary value problem for the Laplace
operator considered in compound media.
%%%%%%%%%%%%%%%%%%%%%%%%%%%%%%%%%%%%%%%%%%%%%%%%%%%%%%%%%%%%%%%%%%%%%%%%%%%%%%%%%%%%%%%%%%%%%%%%%%%%%%%%%%%%%%%%%%%

\section{Heat potentials}
\label{HP} In order to recall the basic facts from the potential
theory, we first address the Laplace equation
\begin{equation}
\label{2-1}
\Delta \,u=0
\end{equation}
considered in the $d$-dimensional Euclidean space $\Bbb R_d$, which is divided
by a smooth closed surface $S$ into a compact internal domain $D_+$ and
external one $D_-$. On the surface $S$ the relevant boundary conditions should
be imposed, that depend on the physical content of the problem in question. For
example, it may be a mathematical formulation of the electrostatic
problem.\cite{Stratton}
 The surface $S$  is supposed
 to posses the properties of smoothness needed. In the potential
theory\cite{Gunter,Kellogg,Smirnov,RR} this  implies that $S$ is the Lyapunov
type surface. The points of Euclidean space $\Bbb R_d$ are denoted by
$x,y,z,\dots$, and $r_{xy}$ is the Euclidean distance between $x$ and $y$. At
any point $x$ on $S$ there exists a unit normal $n_x$ or $n(x)$. For
definiteness we chose the inward directed normal. It is possible because we are
dealing with  closed surfaces~$S$.

For a linear homogeneous  differential equation the fundamental (or elementary) solution
is defined, which is the Green function of this equation in an unbounded space. By definition, the fundamental solution
 obeys the initial equation with a $\delta$-like source on its right-hand side. In the case
of the Laplace equation in $\Bbb R_d$ the fundamental solution is
\begin{equation}
\label{2-2}
E_d(x;x')=-\frac{\Gamma (d/2)\,(r_{xx'})^{-d+2}}{(2d-4)\,\pi^{d/2}}, \quad d\ge3, \quad E_3(x;x')=- \frac{1}{4\pi\, r_{xx'}}
\end{equation}
with the properties
\begin{equation}
\label{2-2a}
\Delta _x\,E_d(x;x')=\Delta_{x'} \,E_d(x;x')=\delta^{(d)} (x,x'),\quad E_d(x;x')=E_d(x';x)\,{.}
\end{equation}

The potentials for the Laplace equation are constructed
by making use of the fundamental solution, namely, the volume potential
\begin{equation}
\label{2-3} U(x)=-\frac{1}{4\pi}\int_D\frac{w(y)}{r_{xy}}\,dy\,{,}
\end{equation}
the single-layer potential
\begin{equation}
\label{2-4}
V(x)=-\frac{1}{4\pi}\int_S\frac{\nu(y)}{r_{xy}}\,dS_y\,{,}
\end{equation}
and the potential of a double layer
\begin{equation}
\label{2-5}
W(x)=-\frac{1}{4\pi}\int_S\mu(y)\frac{\partial }{\partial n_y}\frac{1}{r_{xy}\,}dS_y\,{.}
\end{equation}
These formulas are written for $d=3$, and the following notations are
used: $dy$ and $dS_y$ are, respectively, the elements of the volume
and of the surface at the point $y$, $w(y),\;\; \mu (y)$, and
$\nu(y)$ are the densities of these potentials. It is convenient to
consider the densities $w(y)$, $\mu(y)$, and $\nu(y)$ to be
continuous functions.

All three potentials are  solutions of the Laplace equation (\ref{2-1}),
namely, the volume potential (\ref{2-3}) is harmonic outside $D$, $V(x)$ and
$W(x)$ are harmonic outside $S$ .The single-layer potential $V(x)$ is
continuous everywhere in $\Bbb R_d$, specifically, on passing through $S$. The
potential of a double layer $W(x)$ has a discontinuity on $S$, namely:
\begin{eqnarray}\label{2-6}
W_i(x)&=& W(x) -\frac{1}{2}\,\mu(x)\,{,}\nonumber \\
W_e(x)&=& W(x) +\frac{1}{2}\,\mu(x)\,{,}\quad x\in S\,{.}
\end{eqnarray}
Here $W(x)$ is the value of the integral (\ref{2-5}), when the point $x$  belongs to $S$ ($W(x)$ is a continuous function
for  $x$ varying along $S$), $W_i(x)$  is the value of the double-layer potential (\ref{2-5}), when the point $x$ tends to $S$ from $D_+$, and $W_e(x)$ is the same
when the point $x$ approaches at $S$ from  $D_-$.

 In what follows  we shall frequently use the derivative along the normal to the surface $S$ at the point $y$,
which belongs to $S$. This derivative acts on the function the argument of
which is the distance $r_{xy}$ between
the points $x$ and $y$,  the point $x$  being  not obliged to lay on  $S$.
Simple calculation gives
\begin{equation}
\label{nd} \frac{\partial}{\partial n_y}\,r_{xy} =\cos \varphi\,{,}
\end{equation}
where $\varphi $ is the angle between the vector ${\bf r}_{xy}$, which  starts
at $x$ and ends at $y$,  and the normal ${\bf n}_y $. In the same way we have
\begin{equation}
\label{ndf} \frac{\partial}{\partial n_y}\,f(r_{xy}) = f'(r_{xy})\, \cos
\varphi\,{,} \quad \frac{\partial}{\partial n_y}\,\left ( \frac{1}{r_{xy}}
\right ) = - \frac{\cos \varphi}{r^2_{xy}}\,{.}
\end{equation}

Equations (\ref{nd}) and (\ref{ndf}) are obviously valid for both inward and
outward directed normals.

The application of the potentials (\ref{2-3}) -- (\ref{2-6}) for transforming
the boundary problems for the Laplace equation (\ref{2-1}) to the integral
equations can be found in many textbooks on mathematical
physics.\cite{RR,Smirnov,Kellogg,JS,Roach,Barton,Kupradze}

Let us proceed to consideration of the heat-conduction equation
\begin{equation}
\label{2-7}
\frac{\partial u}{\partial t}-a^2\Delta u=0\,{.}
\end{equation}
The fundamental (or elementary) solution to this equation in $\Bbb R_d$ is
\begin{equation}
\label{2-8}
E_d(x,t;x',t')=\theta (t-t')\,K_0(x,t;x',t')\,{,}
\end{equation}
where
\begin{equation}
\label{2-9}
K_0(x,t;x',t')=\frac{1}{\left (2a\sqrt{\pi (t-t')}\right )^d}\exp\left[
-\frac{r^2_{xx'}}{4a^2(t-t')}
\right]{.}
\end{equation}
The function $K_0(x,t;x',t')$ (the propagator\cite{Barton} or the heat kernel)
obeys the homogeneous heat equations
\begin{eqnarray}
\left (\frac{\partial }{\partial t}- a^2\Delta_x
\right )K_0(x,t;x',t')&=&0\,{,}\nonumber \\
\left (\frac{\partial }{\partial t'}+ a^2\Delta_{x'}
\right )K_0(x,t;x',t')&=&0\,{,}
\end{eqnarray}
and inhomogeneous initial condition
\begin{equation}
\label{2-11}
\ko \to \delta^{(d)}(x-x'),\quad \text{when} \quad t\to t'\,{.}
\end{equation}
This condition enables one to construct the solution of the Cauchy problem for
the nonhomogeneous heat equation
\begin{equation}
\label{2-14pr} \left ( \frac{\partial }{\partial t}- a^2 \Delta_x
 \right ) u(x,t) = f(x,t)
\end{equation}
considered in an unbounded space
\begin{equation}
\label{2-12}
u(x,t)= \int dx'\,\ko \,u_0(x')+
\int_{t'}^{t}d\theta\int dx'\,K_0(x,t;x',\theta )\,f(x',\theta)\,{,}
\end{equation}
where $u_0(x)=u(x,t=t')$. In the classical mathematical
physics\cite{RR} the representation (\ref{2-12}) of the solution to
the heat conduction equation (\ref{2-14pr}) is known as the Poisson
formula. The last term in (\ref{2-12}) can be considered as an analog
of Eq.\ (\ref{2-3}) defining the volume potential for the Laplace
equation.

In a complete agreement with the definitions (\ref{2-4}) and (\ref{2-5})
the heat surface potentials are introduced, namely, the simple-layer potential
\begin{equation}
\label{2-13}
V(x,t)=a^2\dto    K_0(x,t;y, \theta) \,\nu(y, \theta)
\end{equation}
and the potential of a double layer
\begin{eqnarray}
\label{2-14}
 W(x,t)&=&a^2\dto \frac{\partial K_0}{\partial
n_y}(x,t;y,\theta)\,\mu (y,\theta)\nonumber \\
&=&-\int_0^t\frac{d\theta}{2(t-\theta)}\int_S dS_y\,r_{xy}\cos \varphi
\,K_0(x,t;y,\theta \,\mu (y,\theta)\,{,}
\end{eqnarray}
where the functions $\nu(y,\theta)$ and $\mu (y,\theta)$ are the
surface densities of these potentials. By the construction the heat
potentials $V(x,t)$ and $W(x,t)$ vanish at $t=0$.

  For bounded density $\nu(x,t)$ the heat potential of a single layer $V(x,t)$ is continuous everywhere in
   $\Bbb R_d$, also on passing through the surface $S$,
and satisfies the homogeneous heat-conduction equation (\ref{2-7}) outside $S$,
i.e., it is parabolic outside $S$. The normal
 derivatives of $V(x,t)$ have jump discontinuities on $S$. For continuous in $S$ density $\nu(x,t)$  these discontinuities are given by
\begin{eqnarray}
\left (
\frac{\partial V(x,t)}{\partial n_x}
\right )_i&=&\frac{\partial V(x,t)}{\partial n_x}-\frac{1}{2}\,\nu(x,t)\,{,} \nonumber \\
\left (
\frac{\partial V(x,t)}{\partial n_x}
\right )_e&=&\frac{\partial V(x,t)}{\partial n_x}+\frac{1}{2}\,\nu(x,t){,}\quad x \in S\,{.}
\end{eqnarray}

  For bounded density $\mu(x,t)$ the heat potential of a double layer (\ref{2-14}) is continuous
everywhere outside of $S$ (in $\Bbb R_d\backslash S$) and in $S$. Outside $S$
the potential $W(x,t)$ is parabolic. On passing through $S$ it has
discontinuities. For continuous in $S$ density $\mu (x,t)$ these
discontinuities are given by ($d=3$)
\begin{eqnarray}
W_i(x,t)&=&W(x,t) +\frac{1}{2}\,\mu(x,t), \nonumber \\ \label{2-15}
W_e(x,t)&=&W(x,t)-\frac{1}{2}\,\mu(x,t), \quad x\in S.
\end{eqnarray}
The normal derivatives of the double-layer potential are continuous  on passing
through~$S$.

The employment of the Newtonian and heat potentials for transforming the
boundary-value problems for Laplace and heat equations into the integral ones
is based on the discontinuity properties  on the boundary of the double layer
potential and the normal derivatives of the single layer potential. Let us
consider a simple example, namely, construction of the solution to the
Dirichlet problem for the heat equation  in a compact domain $D$ bounded by a
smooth surface $S$:
\begin{eqnarray}
\label{2-16} \left ( \frac{\partial}{\partial t} - a^2 \Delta
\right ) u(x,t)&=&0, \quad x\in D\quad t>0\,{,}\\
u(x,0)&=& 0, \quad x\in D\,{,} \nonumber \\
   u(x,t)&=&\psi (x,t), \quad
x\in S,\quad t>0\,{,} \nonumber
\end{eqnarray}
where the function $\psi (x,t)$ specifies the temperature on the boundary $S$
at different time instants $t$. We shall look for the solution $u(x,t)$ in
terms of the heat potential of a double layer (\ref{2-14})
\begin{equation}
\label{2-18}
u(x,t)= W(x,t)\,{.}
\end{equation}
With account of Eq.\ (\ref{2-15}) we have on the boundary $S$
\begin{equation}
\label{2-19} \frac{1}{2}\, \mu (x,t) =-a^2
\int_0^tdt'\int_SdS_{x'}\frac{\partial K_0}{\partial
n_{x'}}(x,t;x',t')\,\mu(x',t')+\psi(x,t){,} \quad x,x' \in  S\,{.}
\end{equation}
Thus, the problem under consideration is reduced to the solution of the linear integral equation of the second kind.
With respect to the variable $t$  these equations are of the Volterra type and with respect to the spatial variables $x$ and $x'$ they are of
the Fredholm type, the variables $x$ and $x'$ ranging on the boundary $S$.

In an analogous way the integral equations for the Green's function  of the
heat equation can be deduced. Let us consider this technique in the case of the
first boundary-value problem (the Dirichlet problem) for this equation. The
Green function\footnote{Strictly speaking, the function $K(x,t;x',t')$ is the
propagator    of the heat equation. It should be multiplied by the step
function $\theta (t-t')$ in order to get the Green
function\protect.\cite{Barton} This point should be kept in mind when dealing
with the parabolic equations.}
 $K(x,t;x',t')$ is specified by the following conditions:\cite{RR}  it should satisfy
the homogeneous heat equation with respect to the first pair of its arguments
\begin{equation}
\label{2-20}
\left (
\frac{\partial}{\partial t} - a^2 \Delta _x
\right ) K(x,t;x',t')=0\,{,}
\end{equation}
it should obey the inhomogeneous initial condition
\begin{equation}
\label{2-21}
 K(x,t;x',t)=\delta (x,x'), \quad t \ge 0\,{,}
\end{equation}
and the  homogeneous boundary condition
\begin{equation}
\label{2-22}
 K(x,t;x',t')=0, \quad x\in S\,{.}
\end{equation}

We represent the Green function $K(x,t;x',t')$ as the sum of a free propagator and the heat potential of a double layer
\begin{equation}
\label{2-23}
K(x,t;x',t') =K_0(x,t;x',t')+a^2 \dt \frac{\partial K_0}{\partial n_y}(x,t;y,\theta)\, \mu(y,\theta; x',t')\,{.}
\end{equation}
The right-hand side of this equation obviously satisfies Eq.\ (\ref{2-20}). When $t=t'$ the double layer potential in (\ref{2-23})
(the second term) vanishes. The free propagator in this formula $K_0$ enables one to obey the initial condition (\ref{2-21}).
 The density of the double layer potential $\mu (y,\theta; x',t')$ is determined from the boundary condition (\ref{2-22}).
On substituting Eq.\ (\ref{2-23}) into (\ref{2-22}) the following integral equation is obtained for the potential density $\mu$
\begin{equation}
\label{2-24}
\frac{1}{2} \,\mu (x,t; x',t') =- K_0(x,t;x',t') -a^2\dt \frac{\partial K_0}{\partial n_y}(x,t;y,\theta)\,
\mu (y,\theta;x',t'),\quad x\in S\,{.}
\end{equation}

 For Eq.\ (\ref{2-24}) and consequently for Eq.\ (\ref{2-23}) the method of successive approximations can be developed, as Eq.\ (\ref{2-24})
is an integral equation of the second kind. It has been proved, that the series
arising here is uniformly convergent.\cite{Tikhonov}  The first terms of this
series for Eq.\ (\ref{2-23})  are
\[
K(x,t;x',t')=K_0(x,t;x',t')+(-2a^2)^1 \dt \frac{\partial K_0}{\partial
n_y}(x,t;y,\theta)\, K_0(y,\theta; x',t')
\]
\begin{equation}
\label{2-25}
+(-2a^2)^2\dt  \frac{\partial K_0}{\partial n_y}(x,t;y,\theta)\int_{t'}^\theta d\theta_1 \int  dS_{y_1} \frac{\partial K_0}{\partial n_{y_1}}(y,\theta;y_1,\theta_1)\,
K_0(y_1,\theta_1;x',t')+\ldots \,{.}
\end{equation}
Obviously, this series is a result of successive approximations applied to the integral equation for the complete propagator
\begin{equation}
\label{2-25a}
K(x,t;x',t')=K_0(x,t;x',t')-2a^2 \dt \frac{\partial K_0}{\partial n_y}(x,t;y,\theta)\, K(y,\theta; x',t')\,{.}
\end{equation}

The series (\ref{2-25}) is nothing else as the multiple scattering
expansion for the heat kernel in the problems under
consideration.\cite{BVFS,BFSV} Thus we have derived this expansion
in a rigorous way.

By making use of the Laplace transform in Eq.\ (\ref{2-25})  one can remove the integrations over the
intermediate time variables $\theta$'s
\begin{eqnarray}
\bar K(x,x';p) &=& \bar{K}_0(r_{xx'};p) +(-2a^2)^1\int_S dS_y\, \bar K_1(r_{xy};p)
\,\bar K_0(r_{yx'};p) \nonumber \\
&& +(-2a^2)^2\int_S dS_y \,\bar K_1(r_{xy};p)\int _S dS_{y_1}\, \bar K_1
(r_{yy_1};p)\, \bar K_0(r_{y_1x'};p) +\ldots \,{,}\label{2-30}
\end{eqnarray}
where
\begin{eqnarray}
\bar K(x,x';p) &=&\int^\infty_0 e^{-pt} K(x,t;x',0)\,dt\,{,}\label{2-31} \\
\bar K_0(r_{xy};p) &=&\int^\infty_0 e^{-pt} K_0(x,t;y,0)\,dt =\frac{1}{2\pi
a^2}\, K_0\left (
\frac{r_{xy}}{a}\sqrt p\right )\,{,} \label{2-32} \\
\bar K_1(r_{xy};p) &=&\int^\infty_0 e^{-pt}\, \frac{\partial K_0}{\partial
n_y}(x,t;y,0)\,dt=-\frac{\cos \varphi}{2\pi a^3}\,\sqrt p\, K_1\left (
\frac{r_{xy}}{a}\sqrt p\right ){,}\quad y\in S\,{.} \label{2-33}
\end{eqnarray}
In Eq.\ (\ref{2-33}) $\varphi$ is the angle between the vectors ${\bf r}_{xy}$ and ${\bf n}_y$.
 The Laplace transforms $\bar K_0$ and $\bar K_1$ are calculated for $d=2$.
  They are expressed in terms of
the modified Bessel functions\cite{GR} $K_0(z)$ and  $K_1(z)$. We hope that our
notations will not lead to confusion because the free propagator
$K_0(x,t;x',t')$ and the Bessel function $K_0(z)$ have different number of the
arguments. The series (\ref{2-30}) is the perturbative solution to the
following integral equation
\begin{equation}
\label{2-35} \bar K(x,x';p) = \bar K_0(r_{xx'};p) -2a^2\int_S dS_y\, \bar
K_1(r_{xy};p)\,\bar K(r_{yx'};p)\,{.}
\end{equation}
The series (\ref{2-25}) or (\ref{2-30}) contains complete information about the
Green function (heat kernel) in the problem at hand. However, extracting it
from here is not a simple task.

   The second term at the right hand side of Eq.\ (\ref{2-30}) is responsible
for one `reflection' from the boundary (the Born approximation). Its
contribution into the heat kernel can be expressed in terms of the confluent
hypergeometric function\cite{GR} $W_{\alpha \beta}$. By making use of the
convolution theorem for the Laplace transform\cite{Carslaw,CE,GR,LSh} we obtain
\begin{equation}
\label{2-36} \bar K^{(1)}(x,x';p)=\frac{2a^2\sqrt p}{(2\pi)^2a^5}\int_SdS_y\cos
\varphi\,K_0\left (\frac{r_{xy}}{a} \sqrt p\right )\,K_1\left
(\frac{r_{yx'}}{a} \sqrt p\right ){.}
\end{equation}
The inverse Laplace transform gives\cite{Erdelyi}
\begin{equation}
\label{2-37} K^{(1)}(x,x;t)=\frac{1}{2^{5/2}\pi^{3/2}a^2 t}\int_SdS_y\frac{\cos
\varphi}{r_{xy}}\,\exp{\left (-\frac{r^2_{xy}}{2\,a^2t}\right ) }
W_{\frac{1}{2}\frac{1}{2}} \left( \frac{r^2_{xy}}{a^2t} \right){,}
\end{equation}
where $\varphi$ is the angle between the vector ${\bf r}_{xy}$ and the inward
directed normal to the boundary $S$ at the point $y$.

The perturbation series (\ref{2-30}) can be employed, for example, to
find the asymptotic expansion of the heat kernel trace.
% In the next section we shall use this series  for determining
% the coefficients of  the asymptotic expansion of the heat kernel for small
% values of the difference $t-t'$.

%%%%%%%%%%%%%%%%%%%%%%%%%%%%%%%%%%%%%%%%%%%%%%%%%%%%%%%%%%%%%%%%%%%%%%%%%%%%%%%%%%%%%%%%%%%%%%%%%%%%%%%%%%%%%%%%%%%%%%%%%%%%%%%%%%%%%%%%%%

%\section{The heat kernel coefficients from integral equations: compact region}

\section{Compound media}
\label{CM}
 An important advantage of the heat potential technique for
constructing the integral equations is the possibility of  applying
it to compound media. We show this by considering first the solution
of the heat-conduction
 equation instead of the relevant Green's function.

 Thus, in both the regions $D_+$ and $D_-$ the heat equations are defined
\begin{eqnarray}
\left ( \frac{\partial }{\partial t} -a_+^2 \Delta
\right ) u_+(x,t) &\equiv& \hat T_{tx}(a_+) \,u_+(x,t)=0, \quad x\in D_+\,{,}
\label{4-1}\\
\left ( \frac{\partial }{\partial t} -a_-^2 \Delta
\right ) u_-(x,t) &\equiv& \hat T_{tx}(a_-) \,u_-(x,t)=0, \quad x\in D_-
\label{4-2}
\end{eqnarray}
with the matching conditions at the interface $S$, namely, when crossing $S$ the  following quantities should be continuous:
temperature
\begin{equation}
\label{4-3}
u_+(x,t)=u_-(x,t), \quad x\in S
\end{equation}
and heat current
\begin{equation}
\label{4-4}
\lambda_+\frac{\partial u_+(x,t)}{\partial n_+(x)}+\lambda_-\frac{\partial u_-(x,t)}{\partial n_-(x)}=0, \quad x \in S\,{,}
\end{equation}
where $n_+(x)$ and $n_-(x)$ are inward normals to the surface $S$ at the point
$x$ for the regions $D_+$ and $D_-$, respectively. These matching conditions
imply, in particular, that there are no heat sources on $S$. The parameters
$a_+,\;a_-,\; \lambda _+$, and $\lambda_-$ specify the material characteristics
of the media.

We shall look for the solution to this problem in terms of the heat potentials
of single layer and double layer. Here the following feature proves to be
important. If the solution $u_+(x,t)$ in the internal region $D_+$ is
represented as the heat potential of a single layer
\begin{equation}
\label{4-5}
u_+(x,t) =a^2_{+}\int_0^t\!\! dt' \int_S\!\! dS_{x'} K^{(+)}_0(x,t;x',t')\,\nu (x',t')\,{,}
\end{equation}
then the solution $u_-(x,t)$ in the external region $D_-$ should be looked for in terms of the heat potential of a double layer
\begin{equation}
\label{4-6}
u_-(x,t) =a^2_{-}\int_0^t\!\! dt' \int_S\!\! dS_{x'} \frac{\partial K^{(-)}_0}{\partial n_-(x')}\,(x,t;x',t')\,\mu (x',t')\,{,}
\end{equation}
where $K_0^{(+)}$ and  $K_0^{(-)}$
are the fundamental solutions of the heat equations (\ref{4-1}) and (\ref{4-2}), which are defined by
the formula (\ref{2-9}) with $a=a_{+}$ and $a=a_{-}$, respectively.

Substituting Eqs.\ (\ref{4-5}) and (\ref{4-6}) into the first matching
condition (\ref{4-3}) we obtain
\[
a^2_{+}\int_0^t dt' \int _S dS_{x'}K_0^{(+)}(x,t;x',t')\,\nu(x',t')
\]
\begin{equation}
\label{4-7}
=\frac{1}{2}\,\mu(x,t)+
a^2_{-}\int_0^t dt' \int _S dS_{x'}
 \frac{\partial K_0^{(-)}}{\partial n_-(x')}\, (x,t;x',t')\,\mu (x',t')\,{,} \quad
x,\,x'\in S\,{.}
\end{equation}
The second matching condition (\ref{4-4}) results in another integral equation
\[
\lambda_{+}a^2_{+} \int_0^t dt' \int _S dS_{x'} \frac{\partial
K_0^{(+)}}{\partial n_+(x)}\, (x,t;x',t')
\,\nu (x',t')-\frac{1}{2}\, \lambda _+
\nu (x,t)
\]
\begin{equation}
\label{4-8}
+\lambda_-a^2_{-} \int_0^t dt' \int _S dS_{x'} \frac{\partial^2 K_0^{(+)}}{\partial n_-(x)\partial n_-(x')}\, (x,t;x'.t')\,\mu (x',t')=0,\quad x,x'\in S\,{.}
\end{equation}

  Thus the problem under consideration is reduced to the solution of the system of two linear integral equations
of the second kind (\ref{4-7}) and  (\ref{4-8}). It is worth noting that we have obtained  homogeneous equations, because
there are no any heat sources in the problem under study. Hence, we are dealing here with the eigenfunctions only.

  Let us proceed to the Green's function $K(x,t;x',t')$ in this problem. In what follows, it
is convenient to represent this function in terms of the following four components
depending on the range of the arguments:
\begin{equation}
\label{4-9}
K(x,t;x',t')=
\cases{
K_{++}(x,t;x',t'),\quad  x,\,x'\in D_+\,{,}\cr
K_{+-}(x,t;x',t'), \quad  x \in D_+,\,x'\in D_-\,{,}\cr
K_{-+}(x,t;x',t'),\quad  x \in D_-, x'\in D_+\,{,}\cr
K_{--}(x,t;x',t'), \quad x,\,x'\in D_{-}\,{.}\cr}
\end{equation}

  The conditions, which specify  the Green function $K(x,t;x',t')$, can be found in the
following way. This function should provide the solution  $\bar u(x,t)$ of the
inhomogeneous boundary-value problem (\ref{4-1}), (\ref{4-2}), (\ref{4-3}), and
(\ref{4-4}) with the heat source $f(x,t)$ in the form
\begin{equation}
\label{4-15} \bar u(x,t) = -\theta (t-t')\int_{t'}^t d\tau \int _{\Bbb R_d}
K(x,t;\xi  ,\tau)\, f(\xi,\tau)\,d\xi \,{,}
\end{equation}
where $\theta (t-t')$ is the step function. The function $\bar u(x,t)$ will satisfy the inhomogeneous heat-conduction equation
\begin{equation}
\label{4-9a}
f(x,t) =\cases {\hat T_{tx}(a_+)\,\bar u(x,t),\quad x\in D_+, \cr
\hat T_{tx}(a_-)\,\bar u(x,t), \quad x\in D_-{,}}
\end{equation}
if  the Green function $K(x,t;x',t')$ obeys the corresponding  homogeneous heat equations  with respect to the first pair of its arguments
\begin{eqnarray}
\hat T_{tx}(a_+) \,K(x,t;x',t')&=&0,\quad x\in D_+\,{,} \nonumber \\
\hat T_{tx}(a_-) \,K(x,t;x',t')&=&0,\quad x\in D_-\,{} \label{4-10}
\end{eqnarray}
and the inhomogeneous initial condition (\ref{2-21}) with respect to the both pairs of its arguments.
In terms of the components (\ref{4-9}) the initial condition (\ref{2-21}) acquires the form
\begin{eqnarray}
K_{++} (x,t;x',t)&=& \delta (x,x'),\quad x, x' \in D_+\,{,}  \label{4-12} \\
K_{--} (x,t;x',t)&=& \delta (x,x'),\quad x, x' \in D_-\,{,} \label{4-13} \\
K_{+-}(x,t;x',t) &=& K_{-+}(x,t;x',t) =0\,{.}
\label{4-14}
\end{eqnarray}

On the right-hand side of the initial conditions (\ref{4-14}) the delta
function $\delta (x,x')$ with $x,x'\in S$ is absent. Thus  in our consideration
we eliminate the treatment of heat sources at the interface. The point is such
sources alter the matching conditions instead of the initial conditions
(\ref{4-4}). In the next section the solution to the heat conduction equation,
defined on a line, will be constructed  for such a configuration by making use
of the heat potential technique.

The matching conditions (\ref{4-3}) and (\ref{4-4}) are directly transformed into the  conditions for the Green function $K(x,t;x',t')$
with respect to the first pair of its arguments
\begin{eqnarray}
K_{++}(x,t;x',t')&=&K_{-+}(x,t;x',t')\,{,}\label{4-16}\\
\lambda_+ \frac{\partial K_{++}}{\partial n_+(x)}(x,t;x',t') &+&\lambda_- \frac{\partial K_{-+}}{\partial n_-(x)}(x,t;x',t') =0\,{,}\label{4-17} \\
K_{+-}(x,t;x',t')&=&K_{--}(x,t;x',t')\,{,}\label{4-18} \\
\lambda_+ \frac{\partial K_{+-}}{\partial n_+(x)}(x,t;x',t') &+&\lambda_- \frac{\partial K_{--}}{\partial n_-(x)}(x,t;x',t' )=0\,{,}\label{4-19}\\
&x\in S\,{.}& \nonumber
\end{eqnarray}
It turns out that the heat equations  (\ref{4-10}), the initial conditions
(\ref{4-12}) -- (\ref{4-14}), and the matching conditions (\ref{4-16}) --
(\ref{4-19}) are enough for construction of the Green function in the case of
compound media in a unique way.

  We shall look for the components of the Green function (\ref{4-9}) in
terms of the heat potentials of a single and double layers with
 respect to the first pair of their arguments, the components $K_{++}$ and $K_{+-}$
 being expressed through the heat potentials of  single layers and the
components $K_{--}$ and $K_{-+}$ through the heat potentials of  double layers.
In order to take into account the inhomogeneous initial conditions (\ref{4-12})
and (\ref{4-13}) for the components $K_{++}$ and $K_{--}$ we add to the chosen
heat potentials (nonsingular part of the Green function) the free propagator
$K^{(+)}_0$ or $K^{(-)}_0$ (singular part of this function)
\begin{eqnarray}
\kpp &=& \kop +a^2_{+}\dt \kopy \,\npp{,} \\
\kpm &=& a_{+}^2\dt \kopy \,\npm\,{,}\\
\kmm  &=&\kom +a^2_{-}\!\!\int^t_{t'}\!\!d\theta \!\!\int_S\!\!dS_y \dkomy \,\nmm {,}  \\
\kmp &=& a^2_{-}\dt \dkomy \,\nmp\,{.}
\end{eqnarray}
The matching conditions (\ref{4-16}) and (\ref{4-17}) give
\[
\kop +a_{+}^2\dt \kopy \, \npp \]
\begin{equation}
\label{4-24} =\frac{1}{2} \nu_{-+}(x,t;x',t') +a_{-}^2\dt\dkomy \,\nmp \,{,}
\end{equation}
\[
\lambda_{+}\frac{\partial K^{(+)}_0}{\partial n_+(x)}(x,t;x',t')+
\lambda_+a_{+}^2\dt \frac{\partial K_0^{(+)}}{\partial n_+(x)}(x,t; y,\theta)\,
\nu_{++}(y,\theta;x',t')
 \]
\[
-\frac{1}{2}\lambda_+ \npp+\lambda_-a_{-}^2 \dt \frac{\partial^2
K_0^{(-)}}{\partial n_-(x)\partial n_-(y)} (x,t;y,\theta)\, \nmp =0\,{,}
\]
\begin{equation}
\label{4-25} x\in S\,{.}
\end{equation}
In the same way we deduce from (\ref{4-18}) and (\ref{4-19})
\[
a_{+}^2\dt \kopy\, \npm
\]
\begin{equation}
\label{4-26} =\kom +\frac{1}{2} \nu_{--}(x,t;x',t') +a_{-}^2 \dt \dkomy
\,\nmm\,{,}
\end{equation}
\[
-\frac{1}{2}\, \lambda _+\nu_{+-}(x,t;x',t') +\lambda_+a_{+}^2\dt \dkop \,\npm
\]
\[
+ \lambda_-a_{-}^2\dt \frac{\partial ^2K^{(-)}_0}{\partial n_-(x)\partial n_-(y)}(x,t;y,\theta)\, \nmm +\lambda_-\dkom =0\,{,}
\]
\begin{equation}
\label{4-27}
x\in S\,{.}
\end{equation}

The sets of integral equations of the second kind (\ref{4-24}), (\ref{4-25})
and (\ref{4-26}), (\ref{4-27}) define the heat kernel for compound media in
full. With respect to spatial variables these equations are of Fredholm type
while regarding time variable they are of Volterra type. It is essential that
the integration over the spatial variables is restricted by the interface $S$
only. Hence the dimension of the initial problem is reduced by~1. By making use
of the Laplace transform one can remove the integration over the time variables
in Eqs.\ (\ref{4-24}) -- (\ref{4-27}) as it has been done in Sec.~II.

Obviously the integral equations for the heat kernel derived here can be also
applied  when the surface $S$ divides the same medium into the regions $D_+$
and $D_+$, i.e. when the constants $a^2_+$ and $a^2_-$ equal.

For constructing the solutions to the integral equations (\ref{4-24}) --
(\ref{4-27}) the perturbation theory can be employed (see Sec.\ II).  The
expansion parameters in this case prove to be the constants $a^2_\pm$ and
$\lambda _\pm$. The perturbation series generated here are nothing else as the
multiple scattering expansion for the heat kernel.\cite{BVFS,BFSV} Thus we have
proposed a rigorous derivation of these expansions both for homogeneous media
and compact regions and for compound media.

% An attempt to apply the heat potential technique,  developed here, for
% obtaining the asymptotic expansion for the heat kernel in the case of compound
% media has been done in Ref.~\onlinecite{Pirozhenko}.

\section{Asymptotic expansion of heat kernel from perturbation series}
\label{AE} In practical applications, especially in QFT, the
asymptotic expansion of the integrated heat kernel when $t\to+0$
proves to be important.\cite{DW,Kirsten,Bordag,Vassilevich} It has
the form
\begin{equation}
\label{n-1} K(t)\equiv \int dx K(x,t;x,0)= (4 \pi
t)^{-d/2}\sum_{n=0,1,2,\ldots}^\infty t ^{n/2}B_{n/2}+ \text{ES}{.}
\end{equation}
In this expansion $d$ is the dimension  of the configuration space
and  ES stands for the exponentially small corrections as $t\to +0$.
We show how to derive this expansion proceeding from the perturbation
series for the integral equations (\ref{4-24}) -- (\ref{4-27}). The
functions $K_{-+}$ and $K_{+-}$ do not contribute to integrated heat
kernel (see subsection \ref{line-c}), thus we have to consider only
$K_{++}$ and $K_{--}$.

For our purposes it is convenient to use such coordinates that in the
vicinity of the surface $S$ the metric is
$g_{ij}\,dx^i\,dx^j=(dx^3)^2+g_{ab}\,dx^a\,dx^b $ where $x^3$ is a
coordinate on the normal to $S$, $x^3$=0 on $S$. In view of the exact
form of the free propagator (\ref{2-9}), one can infer that in each
term of the perturbation series for integral equations (\ref{4-24})
-- (\ref{4-27}) the power in $t$ contributions are given only at the
following conditions: when evaluating the heat kernel trace, the
integration over $dx$ should be spread over the region immediately
adjacent to the boundary $S$ and in the course of the multiple
integration over the boundary $S$ the respective distances $r_{yy'}$
should be also small. Therefore in the vicinity of $S$ we may replace
the squared distance $(x-z)^2$ by several terms of its expansion in
powers of the corresponding geodesic distance $\sigma$ on the surface
$S$
\begin{eqnarray}
(x-z)^2&=&(x_3-z_3)^2+\sigma^2 \left\{1-(x_3+y_3) k_1+
x_3\,z_3 (k_1^2+k_2^2)\right\} \nonumber\\
&&+\sigma^3\left\{-\frac{1}{3}(2 z_3+x_3)k_1'+ x_3 z_3 (k_1 k_1'+k_2
k_2') \right\}+\dots
\label{eq:dist}\\
k_1=L_{ab}\xi^a\xi^b, && k_2=\frac{1}{2}(\varepsilon_{ac}L^{c}_b+
\varepsilon_{bc}L^{c}_a)\,\xi^a \xi^b,\quad
\varepsilon_{ac}=-\varepsilon_{ca}, \quad k_1'\equiv\frac{d
k_1}{d\sigma},\;\;k_2'\equiv\frac{d k_2}{d\sigma}.\nonumber
\end{eqnarray}

The surface area element is
$dS_z=\left(1-\frac{1}{12}R_{ab}\xi^a\xi^b\,\sigma^2+\dots\right)\sigma\,
d\sigma \,d\Omega,$ $\Omega$ parameterizes a unit sphere, $L_{ab}$ is
the second fundamental form on $S$, $R_{ab}$ is intrinsic Ricci
curvature, $\xi$ is a unit tangent vector at $x$ to the geodesics
with the length $\sigma$ joining $z$ to $x$ on $S$ (see, for example,
Ref.~\onlinecite{Moss}).

Here we present the first three terms of the perturbation series
under consideration when $t\to+0\;\;(d=3)$
\begin{eqnarray}
\label{final}
K^{(0)}(t)&=&K^{(0)}_{++}(t)+K^{(0)}_{--}(t)=\frac{t^{-3/2}}{(4\pi
a_+^2)^{3/2}}\,D_+ +
\frac{t^{-3/2}}{(4\pi a_-^2)^{3/2}}\,D_-, \nonumber\\
K^{(1)}_{++}(t)&=&\frac{t^{-1}\,S}{8 \pi a_+^2}+
\frac{t^{-1/2}}{8\pi^{3/2}a_+}\int_{S}dS\,L_{a}^a+
\frac{t^0}{2^8\,\pi }\int_{S}dS
\left[5\,(L_{a}^{a})^2+L_{a}^{b}L_{b}^{a}- \frac{2}{3}R^a_a\right ]
+...,\nonumber\\
K^{(2)}_{++}(t)&=&-\frac{ t^{-1}}{8
\pi}\,\frac{\lambda_-}{\lambda_+}\, \frac{S}{a_{+}^2}-\frac{1}{8
{\pi}^{3/2}}\frac{\lambda_-}{\lambda_+} \frac{t^{-1/2}}{a_+}
\int_{S}dS\,L_{a}^a\nonumber\\
&&+\frac{t^0}{32 \,\pi}\Biggl\{ a_-\,
\frac{\lambda_-}{\lambda_+}\frac{(a_++2\,a_-)}{(a_++a_-)^2}
\int_{S}dS\left [-(L_{a}^{a})^2+4\,L_{a}^{b}L_{b}^{a}-\frac{1}{3}R_a^a\right ] \nonumber\\
&&+ \Biggl[\frac{1}{8}-
\frac{\lambda_-}{\lambda_+}\frac{1}{(a_++a_-)^3}\left(\frac{35}{12}
a_-^3+
\frac{11}{4}\,a_-^2\,a_+ +2 a_+^2\,a_- \right .\nonumber\\
&&\left .+\frac{2}{3}\,a_+^3
+\frac{9}{4}\frac{a_-^4}{a_+}+\frac{3}{4}\frac{a_-^5}{a_+^2}\right)\Biggr]
\int_{S}dS\left [(L_{a}^{a})^2+2\,L_{a}^{b}L_{b}^{a}\right
]\Biggl\}+\dots
\end{eqnarray}
To obtain $K_{--}^{(1)}(t)$ and $K_{--}^{(2)}(t)$ one should replace
$a_+ \leftrightarrow a_- $, $\lambda_+\leftrightarrow\lambda_-$,
$L_a^b\rightarrow-L_a^b$. The asymptotics of the subsequent terms of
perturbation series may be found in a similar way.  After that all
factors appearing with the same powers of $t$ are added up to give
the heat kernel coefficients. The latter are expressed through the
integrals of the surface geometric invariants. The asymptotics
(\ref{final}) were presented in Ref.~\onlinecite{Pirozhenko} without
considering the derivation of the relevant integral equations.

\section{Heat kernel on a line}
\label{line} In this section we demonstrate the efficiency of our
approach based on integral equations for constructing the heat kernel
on a line. In this case the interface between the media reduces to a
point. As a result we are dealing with the Volterra integral
equations in respect of one (time) variable. These equations are of a
special type (Abel equations), and their solutions can be found in an
exact form.
\subsection{Homogeneous media with gluing conditions}
 By making use of the heat
potential technique we construct here, in an exact form, the heat kernel
$K(x,y;t)$  for the  Laplace operator on an infinite line for homogeneous
medium with a nonstandard gluing conditions at the origin (these conditions
will be specified below). From the physical point of view $K(x,y;t)$  is the
temperature at the point $x$ which is generated by a unit instantaneous heat
source placed at the point $y$ at the moment $t=0$.

    As in previous sections we first  formulate the conditions that define the heat
kernel in the problem under consideration. With respect to the first argument
$K(x,y;t)$ should satisfy the one-dimensional heat-conduction equation
\begin{equation}
\label{A1} \left (\frac{\partial}{\partial t} -\frac{\partial^2 }{\partial x^2}
\right ) K(x,y;t)=0, \quad t > 0,\; -\infty <x < \infty
\end{equation}
and special conditions at the interface $x=0$
\begin{eqnarray}
l\,K(-0,y;t)& =& l^{-1}K(+0,y;t) \,{,}\label{gc1}\\
\left. l^{-1}\frac{\partial }{\partial x} K(x,y;t) \right | _{x=-0} &=& \left
.l\frac{\partial }{\partial x} K(x,y;t)\right | _{x=+0} \,{,}\label{gc2}
\end{eqnarray}
where $l$ is a dimensionless parameter. We shall refer to these conditions as
to gluing ones. In the Appendix A it is shown that such conditions lead to a
selfadjoint spectral problem for  the Laplace operator in any dimension. The
initial condition for $K(x,y;t)$ involves its both space arguments
\begin{equation}
\label{in_cond} K(x,y;0)=\delta(x,y).
\end{equation}

We shall seek for $K(x,y;t)$  in terms of free heat kernel and single layer
heat potentials.\footnote{In multi-dimensional problems we may also choose
single layer potential for internal region and double layer potential for
external region. However when $d=1$ such a choice may lead to divergent
integrals.} The solution is decomposed in four components related to different
positions of the heat source and the  observer
\begin{eqnarray}
K_{-+}(x,y;t)&=&\int_{0}^{t}d\tau \, K_0(x,0;t-\tau)\,\alpha_1(\tau,y), \quad x<0,\; y>0\,{,}\label{pm}\\
K_{++}(x,y;t)&=&K_0(x,y;t) + \int_{0}^{t}d\tau \,
K_0(x,0;t-\tau)\,\alpha_2(\tau,y),\quad  x,y>0\,{,}
\label{pp}\\
K_{+-}(x,y;t)&=& \int_{0}^{t}d\tau \, K_0(x,0;t-\tau)\,\alpha_3(\tau,y), \quad x>0,\; y<0\,{,}\label{mp}\\
K_{--}(x,y;t)&=&K_0(x,y;t) + \int_{0}^{t}d\tau \,
K_0(x,0;t-\tau)\,\alpha_4(\tau,y), \quad x,y<0\,{,} \label{mm}
\end{eqnarray}
where $K_0(x,y;t)$ is the free heat kernel (propagator) on an infinite line
\begin{equation}
\label{Ko} K_0(x,y;t)= \frac{1}{2\sqrt {\pi t}}\,e^{-\frac{(x-y)^2}{4t}}\,{.}
\end{equation}

First we substitute Eqs.\ (\ref{pm}) and (\ref{pp})  into the gluing conditions
(\ref{gc1}) and (\ref{gc2}). Then we take into account that the single layer
potential changes smoothly across the boundary, while the normal derivative of
this  potential undergoes a jump
\begin{eqnarray}
K_{-+}(-0,y;t)&=&\frac{1}{2 \sqrt \pi}\int_0^t d\tau \frac{\alpha_1(y,\tau)}{\sqrt{t-\tau}}\,{,} \label{b-}\\
K_{++}(+0,y;t)&=&K_0(0,y;t)+\frac{1}{2\sqrt \pi}\int_0^t d\tau
\frac{\alpha_2(y,\tau)}{\sqrt{t-\tau}}\,{,}
\label{b+}\\
\left .\frac{\partial }{\partial x}K_{-+}(x,y;t)\right|_{x=-0}&=
&-\frac{1}{2}\,\alpha_1(y,t)\,{,}\label{db-}\\
\left.\frac{\partial }{\partial
x}K_{++}(x,y;t)\right|_{x=+0}&=&-\frac{1}{2}\,\alpha_2(y,t)+
\left.\frac{\partial }{\partial x}K_0(x,y;t)\right|_{x=0}\,{.} \label{db+}
\end{eqnarray}
Inserting Eqs.\ (\ref{b-}) and (\ref{b+}) into Eq.\ (\ref{gc1}) one obtains the
Abel integral equation
\begin{equation}
\frac{1}{2\sqrt \pi}\int_0^{t} \frac{d \tau }{\sqrt{t-\tau}}\,
[l^2\alpha_1(y,\tau)-\alpha_2(y;\tau)]=
K_0(0,y;t) \label{Abel}\\
\end{equation}
with the solution\footnote{When considering  the gluing conditions the use of
the single layer potentials leads to the exactly solvable (no iterations
needed!) Abel equation only in 1-dimensional case.}
\begin{equation}
\alpha_1(y,t)-l^{-2}\alpha_2(y;t)=\frac{2 }{ l^{2}\sqrt{\pi}}\,
\frac{d}{dt}\int_0^t \frac{K_0(0,y;\tau)}{\sqrt{t-\tau}}\
d\tau=\frac{1}{2\sqrt{\pi}\,l^{2}} \
\frac{y}{t^{3/2}}\exp\left(-\frac{y^2}{4\,t}\right). \label{eq1}
\end{equation}
The substitution of Eqs.\ (\ref{db-}) and ({\ref{db+}}) into  Eq.\ (\ref{gc2})
gives
\begin{equation}
\alpha_1(y,t)=-l^2\,\alpha_2(y,t)+\frac{l^2}{2\sqrt{\pi}}\,\frac{y}{t^{3/2}}
e^{-\textstyle{\frac{y^2}{4 t}}}. \label{eq2}
\end{equation}
From (\ref{eq1}) and (\ref{eq2}) it follows  that
\begin{equation}
\alpha_1(y,t)=\frac{l^2}{l^4+1}\,\frac{y}
{\sqrt{\pi}\,t^{3/2}}\;e^{-{\textstyle\frac{y^2}{4\,t}}},\quad
\alpha_2(y,t)=\frac{l^4-1}{l^4+1}\,\frac{y}
{2\sqrt{\pi}\,t^{3/2}}\;e^{-{\textstyle\frac{y^2}{4\,t}}}.
\end{equation}
And finally
\begin{eqnarray}
K_{-+}(x,y;t)&=&\frac{l^2}{l^4+1}\,\frac{1}{\sqrt{\pi\,t}} \;
e^{-{\textstyle\frac{(x-y)^2}{4t}}}, \label{Res-+}\\
K_{++}(x,y;t)&=&\frac{1}{2\sqrt{\pi\,t}}\;
e^{-{\textstyle\frac{(x-y)^2}{4t}}}+\frac{l^4-1}{l^4+1}\,\frac{1}{2\sqrt{\pi\,t}}
\; e^{-{\textstyle \frac{(x+y)^2}{4t}}}. \label{Res++}
\end{eqnarray}
In a similar way one gets
\begin{eqnarray}
K_{+-}(x,y;t)&=&\frac{l^2}{l^4+1}\,\frac{1}{\sqrt{\pi\,t}} \;
e^{-{\textstyle\frac{(x-y)^2}{4t}}}, \label{Res+-}\\
K_{--}(x,y;t)&=&\frac{1}{2\sqrt{\pi\,t}}\;
e^{-{\textstyle\frac{(x-y)^2}{4t}}}-\frac{l^4-1}{l^4+1}\,\frac{1}{2\sqrt{\pi\,t}}
\; e^{-{\textstyle \frac{(x+y)^2}{4t}}} \label{Res--}
\end{eqnarray}

These formulas are is in a complete agreement with  the result of a combined
employment of the Lemma 5.2 argued in Ref.~\onlinecite{GKV} and Lemma 4.1 from
Ref.~\onlinecite{GKV-LMP}
\begin{eqnarray}
K_{++}(x,y;t)&=&\cos^2 \theta K_N(x,y;t) + \sin^2\theta
K_D(x,y;t)\,{,}\label{pp1}\\
K_{--}(x,y;t)&=&\sin^2 \theta K_N(x,y;t) + \cos^2\theta
K_D(x,y;t)\,{,}\label{mm1}\\
K_{+-}(x,y;t)&=&\sin \theta\, \cos\theta \,[K_N(x,y;t)-K_D(x,y;t)]\,{,}
\label{pm1}
\end{eqnarray}
where
$$\cos^2 \theta=\frac{l^4}{l^4+1},\quad \sin^2 \theta=\frac{1}{l^4+1}\,{,}$$
and $K_D(x,y;t)$ and $K_N(x,y;t)$ are the heat kernels for Dirichlet and
Neumann boundary conditions, respectively.

\subsection{Dielectric-like conditions on a line} We construct here
the heat kernel for  the Laplace operator defined on an infinite line with
dielectric-like matching conditions at the origin $x=0$. The heat kernel is
defined by the heat conduction equation
\begin{equation}
\label{A2} \left ( \frac{\partial }{\partial t}-
a^2(x)\frac{\partial^2}{\partial x^2}\right )  K(x,y;t)=0,\quad t>0,\; -\infty
<x<+\infty,\; x\not= 0\,{,}
\end{equation}
where \[ a^2(x)=\cases{a_{-}^2,\quad & $x<0$, \cr a_{+}^2,\quad &$x>0$,\cr}
\]
 $a_{+}^2$ and $a_{-}^2$ being positive constants. At the interface of
dielectric media the matching conditions
\begin{eqnarray}
K(-0,y;t)& =&K(+0,y;t) \,{,}\label{dc1}\\
\left. \lambda_{-} \frac{\partial }{\partial x}K(x,y;t)\right |_{x=-0}&=&
\left. \lambda_{+} \frac{\partial }{\partial x}K(x,y;t)\right |_{x=+0}
\label{dc2}
\end{eqnarray}
should be met. As usual, the initial condition for $K(x,y;t)$ is given by
(\ref{in_cond}).

We call the boundary conditions (\ref{dc1}) and (\ref{dc2}) the dielectric-like
conditions. The use of this term requires some explanations. When two
dielectric media $D_+$ and $D_-$ possessing different characteristics are
separated by the interface $S$ of an arbitrary form then in Maxwell theory we
have on the surface $S$ the set of coupled boundary conditions involving all
the components of the electromagnetic potential $A_\mu (t, {\bf x}), \quad \mu
=0,\ldots, d$. If one disregards the vector character of the electromagnetic
field and confine oneself to oscillations described by a sole scalar potential
(for example, sound waves which are described by a scalar velocity potential)
then at the interface between different media the conditions (\ref{dc1}) and
(\ref{dc2}) should be satisfied. In other words,  these boundary conditions
hold in the theory of scalar `photons' in compound media.

 Again we seek for the solution in terms of a relevant free
propagator and single layer heat potentials:
\begin{eqnarray}
K_{-+}(x,y;t)=a^2_{-}\int_{0}^{t}d\tau K_0(x,0;
a^2_{-}(t-\tau))\,\beta_1(\tau,y),\;
x<0, y>0\,{,} \label{diel_pm}\\
K_{++}(x,y;t)=K_0(x,y;a_{+}^2 t) + a^2_{+}\int_{0}^{t}\!\!d\tau
K_0(x,0;a_{+}^2(t-\tau))\,\beta_2(\tau,y),\;
x,y>0, \label{diel_pp}\\
K_{+-}(x,y;t)= a^2_{+}\int_{0}^{t}d\tau K_0(x,0;a^2_{+}(t-\tau))\,\beta_3(\tau,y),\; x>0,y<0, \label{diel_mp}\\
K_{--}(x,y;t)= K_0(x,y;a^2_{-}\,t) + a^2_{-}\int_{0}^{t}\!\!d\tau
K_0(x,0;a^2_{-}(t-\tau))\beta_4(\tau,y),\; x,y<0{.} \label{diel_mm}
\end{eqnarray}

First we insert Eqs.\ (\ref{diel_pm}) and  (\ref{diel_pp}) into the matching
conditions (\ref{dc1}) and (\ref{dc2}). The single layer potential changes
smoothly across the boundary, while its  normal derivative  undergoes a jump
\begin{eqnarray}
K_{-+}(-0,y;t)&=&\frac{a_{-}}{2\sqrt \pi}\int_0^t d\tau \frac{\beta_1(y,\tau)}{\sqrt{t-\tau}}\,{,} \label{diel_b-}\\
K_{++}(+0,y;t)&=&K_0(0,y;a^2_{+}t)+\frac{a_{+}}{2\sqrt \pi}\int_0^t d\tau
\frac{\beta_2(y,\tau)}{\sqrt{t-\tau}}\,{,}
\label{diel_b+}\\
\left.\frac{\partial }{\partial x}K_{-+}(x,y;t)\right|_{x=-0}&=
&-\frac{1}{2}\,\beta_1(y,t),\label{diel_db-}\\
\left.\frac{\partial }{\partial
x}K_{++}(x,y;t)\right|_{x=+0}&=&-\frac{1}{2}\,\beta_2(y,t)+
\left.\frac{\partial }{\partial x}K_0(x,y;a_+^2t)\right|_{x=0}\,{.}
\label{diel_db+}
\end{eqnarray}
Substituting Eqs.\  (\ref{diel_b-}) and (\ref{diel_b+})  into Eq.\
(\ref{dc1}) one obtains the Abel integral equation
\begin{equation}
\frac{1}{2\sqrt \pi}\int_0^{t} \frac{d \tau
}{\sqrt{t-\tau}}\,[a_{-}\beta_1(y,\tau)-a_{+}\beta_2(y;\tau)]=
\,K_0(0,y;a^2_{+}t) \label{Abel2}\\
\end{equation}
with the solution
\begin{eqnarray}
a_{-}\beta_1(y,t)-a_{+}\beta_2(y;t)&=&\frac{2}{\sqrt{\pi}}\,
\frac{d}{dt}\int_0^t \frac{d \tau}{\sqrt{t-\tau}}K_0(0,y;a^2_{+}\tau)\nonumber
\\
   &=&\frac{1}{2\sqrt{\pi}} \ \frac{y}{a^2_{+} \
t^{3/2}}\exp\left(-\frac{y^2}{4\,a_{+}^2 t}\right). \label{eq37}
\end{eqnarray}
The substitution of Eqs.\ (\ref{diel_db-}) and ({\ref{diel_db+}}) into Eq.\
(\ref{dc2}) gives
\begin{equation}
\lambda_-
\beta_1(y,t)=-\lambda_+\,\beta_2(y,t)+\lambda_+\frac{1}{2\sqrt{\pi}}\,\frac{y}{a^3_{+}t^{3/2}}
e^{-\textstyle{\frac{y^2}{4 a^2_{+}t}}}. \label{eq38}
\end{equation}
From Eqs.\  (\ref{eq37}) and (\ref{eq38}) it follows  that
\begin{equation}
\beta_1(y,t)=\frac{\frac{\lambda_+}{a_{+}}}{\frac{\lambda_+}{a_{+}}+
\frac{\lambda_-}{a_{-}}}\,\frac{y}
{\sqrt{\pi}\,a_{+}^3\,t^{3/2}}\;e^{-{\textstyle\frac{y^2}{4\,a_{+}^2\,t}}},\quad
\beta_2(y,t)=\frac{\frac{\lambda_+}{a_{+}}-\frac{\lambda_-}{a_{-}}}{\frac{\lambda_+}{a_{+}}+
\frac{\lambda_-}{a_{-}}} \,\frac{y}{2\sqrt{\pi}
a_{+}^3\,t^{3/2}}\;e^{-{\textstyle\frac{y^2}{4\,a_{+}^2 t}}}.
\end{equation}
And finally
\begin{eqnarray}
K_{-+}(x,y;t)&=&\frac{\frac{\lambda_+}{a_{+}}}{\frac{\lambda_+}{a_{+}}+
\frac{\lambda_-}{a_{-}}}\,\frac{1}{\sqrt{\pi\,a_{+}^2\,t}} \;
e^{-{\textstyle\frac{(x-y\,a_{-}/a_{+})^2}{4\,a_{-}^2\,t}}}, \label{diel-+}\\
K_{++}(x,y;t)&=&\frac{1}{2\sqrt{\pi\,a_{+}^2\,t}}\;
e^{-{\textstyle\frac{(x-y)^2}{4\,a_{+}^2\,t}}}+
\frac{\frac{\lambda_+}{a_{+}}-\frac{\lambda_-}{a_{-}}}{\frac{\lambda_+}{a_{+}}+
\frac{\lambda_-}{a_{-}}}\,\frac{1}{2\sqrt{\pi\,a_{+}^2\,t}} \; e^{-{\textstyle
\frac{(x+y)^2}{4\,a_{+}^2\,t}}}. \label{diel++}
\end{eqnarray}
In a similar way one gets
\begin{eqnarray}
K_{+-}(x,y;t)&=&\frac{\frac{\lambda_-}{a_{-}}}{\frac{\lambda_-}{a_{-}}+
\frac{\lambda_+}{a_{+}}}\,\frac{1}{\sqrt{\pi\,a_{-}^2\,t}} \;
e^{-{\textstyle\frac{(x-y\,a_{+}/a_{-})^2}{4\,a_{+}^2\,t}}}, \label{diel+-}\\
K_{--}(x,y;t)&=&\frac{1}{2\sqrt{\pi\,a_{-}^2\,t}}\;
e^{-{\textstyle\frac{(x-y)^2}{4\,a_{-}^2\,t}}}+
\frac{\frac{\lambda_-}{a_{-}}-\frac{\lambda_+}{a_{+}}}{\frac{\lambda_-}{a_{-}}+
\frac{\lambda_+}{a_{+}}}\,\frac{1}{2\sqrt{\pi\,a_{-}^2\,t}} \; e^{-{\textstyle
\frac{(x+y)^2}{4\,a_{-}^2\,t}}}. \label{diel--}
\end{eqnarray}
The solution obtained here exactly reproduces the results obtained in this
problem by other methods.\cite{Sommerfeld,Carslaw}
\subsection{$\delta$-Like heat source at the interface}
\label{line-c} In preceding considerations we excluded the
configuration when the $\delta$-like heat source is placed at the
interface of the media $y=0$. For completeness we have to check
whether this configuration contributes to the trace
$\int_{-\infty}^\infty K(x,x;t)\,dx$. To this end we use the approach
of Ref.\ \onlinecite{Schaaf47}. The idea is to modify the boundary
conditions (\ref{dc1}) and (\ref{dc2}) so that they allow for the
heat source placed at the interface
\begin{equation}
\label{mod_dc1} K_-(-0;t)=K_+(+0;t)\,{,} \end{equation}
\begin{equation}
\label{mod_dc2} \left .\lambda_-\frac{\partial}{\partial x}K_-(x;t)\right
|_{x=-0}- \left .\lambda_+\frac{\partial}{\partial x}K_+(x;t)\right |_{x=+0}=
 Q\, \delta(t) \,,
\end{equation}
The condition (\ref{mod_dc2}) means that the heat $Q$ instantly generated by
the source  is divided into two flows which are proportional to $\lambda_{-}$
 and $\lambda_{+}$. We represent the solution in
the form
\begin{eqnarray}
K_-(x;t)&=&a^2_{-}\int_{0}^{t}d\tau K_0(x,0; a^2_{-}(t-\tau))\,\beta_1(\tau), \quad x<0\,{,}\label{diel+}\\
K_+(x;t)&=& a^2_{+}\int_{0}^{t}d\tau K_0(x,0;a_{+}^2(t-\tau))\,\beta_2(\tau),
\quad x>0\,{.} \label{diel-}
\end{eqnarray}
The functions $\beta_1(\tau)$ and $\beta_2(\tau)$ can be determined by  making
use of the Laplace transform. We denote by $\bar K(x;s)$ the transform of
$K(x;t)$, i.e.
\[
\bar K(x;p)=\int_{0}^{\infty}dt \, e^{-p\,t} \, K(x;t).
\]
Transforming (\ref{diel+}) and (\ref{diel-}) we obtain
\begin{eqnarray}
\bar K_-(x;p)&=&a_{-} \bar\beta_-(p)\sqrt{\pi /p}\,\exp(-x \sqrt{p}/a_{-}) ,
 \label{laplace+}\\
\bar K_+(x;p)&=&a_{+} \bar\beta_+(p)\sqrt{\pi/p}\,\exp(-x \sqrt{p}/a_{+}) .
\label{laplace-}
\end{eqnarray}
Then Eq.\ (\ref{mod_dc1}) leads to the relation between the Laplace transforms
$\bar \beta_-$ and $\bar\beta_+$
\begin{equation}
a_{-}\bar \beta_-(p)-a_{+}\bar \beta_+(p)=0\,{.} \label{rel_1}
\end{equation}
The substitution of Eqs.\ (\ref{diel+}) and (\ref{diel-}) into Eq.\
(\ref{mod_dc2}) gives
\begin{equation}
\lambda_- \beta_-(t)+\lambda_+\,\beta_+(t)=2\,Q\,\delta(t).
\end{equation}
After the Laplace transform one arrives at the second relation between $\bar
\beta_1$ and $\bar \beta_2$
\begin{equation}
\lambda_1 \,\bar \beta_-(p)+\lambda_2\,\beta_+(p)=2\,Q\,\bar\delta(p)\,{.}
\label{rel_2}
\end{equation}

How to apply the Laplace transform to the singular $\delta $ function can be
found in appropriate handbooks.\cite{CE,LSh} The essence of the matter comes to
defining the integration rule
\[
\int_0^\infty f(t)\,\delta (t)\, dt= f(0)\,{,}
\]
whence it follows in particular
\[
\bar \delta (p) =\int^\infty_0e^{-pt}\delta(t)\,dt=1\,{.}
\]
In the problems treated by the integral Laplace transform the semiaxis $t>0$
(or $t>t_0$) is usually considered. Therefore the $\delta$-function should be
defined here in a nonsymmetric way, for example, as the limit when $\varepsilon
\to +0 $ of the function
\[
\delta _\varepsilon(t)=\cases{0,\quad &$t<0,\;t>\varepsilon$,\cr
{1}/{\varepsilon}, \quad &$0<t<\varepsilon$.\cr}
\]

 The solution of the system (\ref{rel_1}), (\ref{rel_2}) is
\begin{equation}
\label{53} \bar
\beta_-(p)=\frac{2\,a_{+}\,Q\,\bar\delta(p)}{\lambda_+\,a_{-}+\lambda_{-}\,a_{+}},\quad
\bar\beta_+(p)=\frac{2\,a_{-}\,Q\,\bar\delta(p)}{\lambda_+\,a_{-}+\lambda_{-}\,a_{+}}.
\end{equation}
By the inverse Laplace transform we find from (\ref{53})
\begin{equation}
\beta_-(t)=\frac{2\,a_{+}\,Q\,\delta(t)}{\lambda_+\,a_{-}+\lambda_{-}\,a_{+}},\quad
\beta_+(t)=\frac{2\,a_{-}\,Q\,\delta(t)}{\lambda_+\,a_{-}+\lambda_{-}\,a_{+}}.
\label{inv_Laplace}
\end{equation}
Having inserted (\ref{inv_Laplace}) into (\ref{diel+}) and (\ref{diel-}) we
derive
\begin{eqnarray}
K_-(x;t)&=& \frac{1}{\sqrt{\pi\, t}}\; e^{-\textstyle{\frac{x^2}{4
a^2_{-}t}}}\,
\frac{Q}{\textstyle{\frac{\lambda_+}{a_{+}}+\frac{\lambda_{-}}{a_{-}}}},
 \quad x < 0,\label{Res+}\\
K_+(x;t)&=& \frac{1}{\sqrt{\pi\, t}}\; e^{-\textstyle{\frac{x^2}{4
a^2_{+}t}}}\,
\frac{Q}{\textstyle{\frac{\lambda_+}{a_{+}}+\frac{\lambda_{-}}{a_{-}}}},\quad
x>0{.} \label{Res-}
\end{eqnarray}
These formulae show that for $t>0$ the heat kernel is finite notwithstanding
the $\delta$-like heat source situated at the interface between two media.
Therefore the neighborhood of the point $x=0$ gives no contribution to the heat
kernel trace $\int_{-\infty}^\infty K(x,x;t)\,dx$.

\section{Conclusion}
\label{Con} For a broad set of boundary conditions the finding of the
heat kernel is reduced to the solution of integral equations defined
on the boundary (or at the interface) of the manifolds. As a result
the dimension of the initial problem is brought down by 1. Remarkably
this technique is applicable to compound media where the standard
methods for the investigation of heat kernel do not work because in
this case the principal part of the elliptic operator in question is
not smooth.

The perturbation series for the integral equations derived are nothing else as
the multiple scattering expansions for the relevant heat kernels.
 Thus
a rigorous derivation of these expansions both for homogeneous media and
compact regions and for compound media has been done.

The efficiency of this approach is convincingly   demonstrated by
constructing, in an exact form, the heat kernel on an infinite line
with diverse matching conditions and by deriving the first terms of
the asymptotic expansion for integrated heat kernel in the case of
three dimensional compound media.

 \acknowledgments
We are indebted to D.~V.~Vassilevich for valuable discussions of problems
investigated in this paper. The work was partially support by the
Heisenberg-Landau Program and by the Russian Foundation for Basic Research
(grant No 03-01-00025).
\appendix
\section{Self-adjointness of boundary value problems for compound media}
\label{appA} Often it is helpful to  use the spectral representation
for the Green's function of the heat conduction equation~(\ref{2-7})
\begin{equation}
\label{a1} K(x,t;x',0)=\sum_k e^{-\omega_k t}f_k(x)f^*_k(x')\,{,}
\end{equation}
where $f_k(x)$ are the eigenfunctions of the spectral problem at hand
\begin{equation}
\label{a2} -a^2\Delta f_k(x)=\omega_kf_k(x), \quad x\in D
\end{equation}
obeying the relevant boundary conditions on $S$. Apparently, the representation
(\ref{a1}) is well defined if the spectral problem (\ref{a2}) is hermitian and
positive definite. In this connection it is worth elucidating the boundary
conditions for compound media that lead to the self-adjoint spectral problem
(see Eqs.\ (\ref{4-1}) -- (\ref{4-4})). In this case we have instead
of~(\ref{a2})
\begin{equation}
\label{a3} -a^2(x)\, \Delta f_k(x)=\omega _k f_k(x)\,{,}
\end{equation}
where
\begin{equation}
\label{a4} a^2(x)=\cases{a_+^2, \quad &$x\in D_+$\,{,}\cr a_-^2, \quad & $x\in
D_-$\,{,}\cr}
\end{equation}
$a_+^2$ and $a_-^2$ being constants. At the interface $S$ the natural modes
$f_k(x)$ obey the dielectric like conditions
\[
f_{k+}(x)=f_{k-}(x)\,{,}
\]
\[
\lambda _+ \frac{\partial f_{k+}(x)}{\partial n_+(x)}+ \lambda _-
\frac{\partial f_{k-}(x)}{\partial n_-(x)}=0\,{,}
\]
\begin{equation}
\label{a5} x\in S\,{.}
\end{equation}
Here we are using the same notations as in Eqs.\ (\ref{4-1}) -- (\ref{4-4}).

Since the outer region $D_-$ is not bounded the spectrum $\omega _k$ is
continuous. For example, we can assume that at the spatial infinity $|{\bf
x}|\to \infty $ the natural modes $f_k(x)$ satisfy the scattering problem
conditions and decrease sufficiently fast.  The explicit form of these
conditions will not be needed below.

 The differential operator in Eq.\ (\ref{a3}) apparently coincides with its
 adjoint
 \begin{equation}
 \label{a6}
[-a^2(x)\, \Delta ]\dagger =\cases{-a_+^2\Delta, \quad&$x\in D_+\,{,}$\cr
-a_-^2\Delta, \quad &$x\in D_-\,{.}$\cr}
 \end{equation}
The boundary value  problem with the operator $-a^2(x)\,\Delta$ will be
self-adjoint when the integral for two sufficiently smooth functions $u(x)$ and
$v(x)$
\begin{equation}
\label{a7} I=\int_{D_+\cup D_-}\!\!dx\, a^2(x)\left ( v\,\Delta\, u - u\,\Delta
\,v\right )
\end{equation}
vanishes. Applying the Green integral formula for the domains $D_+$ and $D_-$
separately we obtain
\begin{equation}
\label{a8} I=\int_S \!dS\,\left [ a_+^2\left ( u_+\frac{\partial v_+}{\partial
n_+ }- v_+\frac{\partial u_+}{\partial n_+ }\right ) + a_-^2\left (
u_-\frac{\partial v_-}{\partial n_- }- v_-\frac{\partial u_-}{\partial n_-
}\right )\right ]{.}
\end{equation}
The functions $u(x)$ and $v(x)$ are assumed to diminish at the infinity in such
a way that the region $|{\bf x}|\to \infty$ does not contribute to the integral
(\ref{a8}). The integral $I$ is equal to zero, for example, for the following
conditions at the interface~$S$
\[
u_+(x)=u_-(x)\,{,}
\]
\begin{equation}
\label{a9} a_+^2\frac{\partial u_+(x)}{\partial n_+(x)}+a_-^2\frac{\partial
u_-(x)}{\partial n_-(x)}=0\,{,}\quad x\in S\,{.}
\end{equation}
The function $v(x)$ should satisfy the same boundary conditions on~$S$. Thus
the dielectric-like conditions (\ref{a5}) lead to self-adjoint boundary value
problem if
\[
\frac{\lambda _+}{a_+^2}=\frac{\lambda _-}{a_-^2}\,{.}
 \]

Of course, conditions (\ref{a9}) do not exhaust all the cases when the boundary
value problem under consideration is self-adjoint. Let the surface $S$ divides
the same medium into the domains $D_+$ and $D_-$, i.e.\ $a_+^2=a_-^2$. We get
self-adjoint spectral problem if impose at the interface $S$ the following
gluing conditions
\[
l\,u_+(x)=l^{-1}u_-(x)\,{,}
\]
\begin{equation}
\label{a10} l^{-1}\frac{\partial u_+(x)}{\partial n_+(x)}+l\,\frac{\partial
u_-(x)}{\partial n_-(x)}=0,\quad x\in S\,{,}
\end{equation}
where $l$ is a dimensionless constant. The one-dimensional version of
this problem has been considered in Sec.~\ref{line}.

\end{document}